\documentclass[aps,pre,preprint,groupedaddress]{revtex4}
\usepackage{graphicx}
\usepackage{amsmath}
\usepackage{bm}
\usepackage{amsbsy}
\usepackage{subfigure}
\usepackage{CJKutf8}
\usepackage{epstopdf}
\usepackage{epsf,epsfig,graphics}
\usepackage{float}
\usepackage{makeidx}
\usepackage{latexsym}
\usepackage{amssymb}
\usepackage{amsfonts}
\usepackage{mathrsfs}
\usepackage{dcolumn}
\usepackage{amsmath}
\usepackage{color}
\usepackage{lineno,hyperref}
\modulolinenumbers[5]

\usepackage{morefloats}
\DeclareMathAlphabet\mathbfcal{OMS}{cmsy}{b}{n}

\begin{document}
\title{Coexistence of distinct non-uniform non-equilibrium steady states in Ehrenfest multi-urn model on a ring}

\author{Chi-Ho Cheng$^a$\footnote{phcch@cc.ncue.edu.tw}\begin{CJK*}{UTF8}{bsmi}(鄭智豪)\end{CJK*} and Pik-Yin Lai$^{b,c}$\footnote{pylai@phy.ncu.edu.tw}\begin{CJK*}{UTF8}{bsmi}(黎璧賢)\end{CJK*}}
\address{$^a$Dept. of Physics, National Changhua University of Education, Changhua 500, Taiwan, R.O.C.}
\address{$^b$Dept. of Physics and Center
for Complex Systems, National Central University, Chung-Li District, Taoyuan City 320, Taiwan, R.O.C.}
\address{$^c$Physics Division, National Center for Theoretical Sciences, Taipei 10617, Taiwan, R.O.C.}

\begin{abstract}
The recently proposed Ehrenfest $M$-urn model with interactions on a ring is considered as a paradigm model which can exhibit a variety of distinct non-equilibrium steady states. Unlike the previous 3-urn model on a ring which consists of a uniform and a non-uniform non-equilibrium steady states, it is found that for even $M\geq 4$, an additional non-equilibrium steady state can coexist with the original ones. Detailed analysis reveals that this new non-equilibrium steady state emerged via a pitchfork bifurcation which cannot occur if $M$ is odd. Properties of this non-equilibrium steady state, such as stability, and steady-state flux are derived analytically for the 4-urn case. The full phase diagram with the phase boundaries is also derived explicitly. The associated thermodynamic stability is also analyzed confirming its stability.  These theoretical results are also explicitly verified by direct Monte Carlo simulations for the 3-urn and 4-urn ring models.

\end{abstract} 


\maketitle
\section{Introduction}
 Starting from the second law of thermodynamics, non-equilibrium   processes have been under active studies for about two hundred years due to their fundamental importance in classical thermodynamics and statistical mechanics\cite{Huangbook}. In contrast to the well-understood
equilibrium cases, non-equilibrium
statistical physics remained challenging for a long period, partly due to the lack
of well-characterized states or principles  such as  free energy  minimization for equilibrium systems.  
The last three decades marked a breakthrough in the
understanding of nonequilibrium statistical physics, especially
in the far from equilibrium and fluctuation dominating
regimes. New physical laws, such as fluctuation theorems\cite{Jarzynski97,Crooks99}
and theoretical techniques such as stochastic thermodynamics\cite{Sekimoto1998,Seifert2012}, proved to be very successful in a broad range of
non-equilibrium processes in small systems in which thermal
fluctuations dominate.

A major signature for non-equilibrium states that differ 
from the equilibrium ones is that some net
fluxes, such as mass, momentum, heat, or probability, are generated  so that detailed balance is broken.
These fluxes can be transient (as in the case of relaxation towards an equilibrium state), steady (a time-independent constant flux as in the case of non-equilibrium steady-states), or time-varying (as in a system under time-dependent external drives or system with autonomous dynamics).
Experimentally,
a non-equilibrium state can be conveniently generated by creating
  concentration gradients (such as
temperature, velocity, or  potential)  to produce some generalized forces to drive  the system. 
When the generalized force is independent of time,
the system can be driven into a nonequilibrium steady state.
(NESS), which is perhaps the simplest and tractable nonequilibrium states [2,3]. For example, the fluctuation theorem was first discovered\cite{Evans93,Gallavotti95} in NESS and then later extended to other non-steady scenarios.
A time-independent  steady-state distribution, albeit non-Boltzmann, exists in NESS, which can often serve as a convenient quantity in quantitative characterizing the non-equilibrium states theoretically, and also can be measured accurately in experiments or simulations if a sufficient measurement duration is allowed (which is often achievable since the system is steady).
In the NESS, entropy is produced at a positive constant rate on average, which is a measure of irreversibility.

Even for non-equilibrium steady
state (NESS), it is difficult to describe nonequilibrium phase
transitions between different NESSs and their relationship to
some microscopic models. The transition between different NESSs is of  interest, both in the foundation of statistical physics and for designing the concept of  engines driven between NESSs. Important theoretical framework and physical laws, such as  steady state thermodynamics\cite{oono,sasa} and Hatano-Sasa equality\cite{hatano01}  for quantifying transitions between initial and final NESSs were established and experimentally verified\cite{trep04}. 

A variety of experimental systems have been set up to explore the NESS systems. Examples include single Brownian particles in a trap moving at constant speed\cite{andrieux,carberry,wang}
or driven by a constant force across a periodic potential\cite{Speck2007,ma15SM}, power fluctuations in a vertically agitated granular gas\cite{feitosa}, in liquid-crystal electroconvection\cite{goldburg}, temperature and
heat flux fluctuations in turbulent convection\cite{ciliberto,shang}, and
fluctuations of entropy production in driven RC-circuits\cite{zon,ciliberto13,chiang16,chiang17}, autonomous Brownian gyrators\cite{chiang17b,lin22} colloidal monolayers suspended near a liquid-solid interface\cite{ma15,ma17}. These experiments provided new insights into the nature of non-equilibrium processes under NESS conditions. To this end, a theoretically tractable system that exhibits a rich variety of NESS behavior would serve as a paradigm model to gain deeper insights into the detailed properties quantitatively in NESSs.

From a historical perspective, the classic
Ehrenfest two-urn non-interacting model\cite{urn1907} was proposed in 1907 to resolve
the microscopic time
reversal  and Poincare recurrence paradoxes\cite{Poincare,Huangbook} (which accompanied the  H-theorem\cite{Hthm}in 1872 on
explaining how a system approaches equilibrium from non-equilibrium and the associated irreversibility) and clarify the relationship between reversible
microscopic dynamics and irreversible macroscopic thermodynamics.
The classic two-urn Ehrenfest model\cite{urn1907} is a system of $N$
particles distributed in two urns. Each particle in an urn
is chosen randomly and put into the other with equal
probability. The Ehrenfest two-urn model is a simple and tractable
model to clearly illustrates the conceptual foundation of
statistical mechanics and the relaxation towards equilibrium. The model was subsequently solved exactly by Kac\cite{kac} and has  often been used to demonstrate the second law of thermodynamics and the approach to equilibrium.

In recent years, a  model  based on the classic non-interacting Ehrenfest two-urn model for non-equilibrium irreversible processes has been  proposed with the introduction of  particle interactions in a physical way\cite{cheng17}.  This two-urn model, although non-trivial, can be solved to some extent and obtain some nice analytical results. This modified Ehrenfest model with particle interaction  explicitly imposed,  opened a new avenue to study various non-trivial nonequilibrium statistical mechanics in an analytically tractable model.  
Such urn model with inter-particle interactions within the same urn has been further generalized to an arbitrary number of urns and the equilibrium properties, such as uniform and non-uniform population states and the associated first-order transition, were sorted out analytically in details\cite{cheng20}. 
 Subsequently, the interacting Ehrenfest model was generalized to study  the non-equilibrium steady-states in the 3-urn system with bias transition probabilities\cite{cheng21}.
The 3-urn system has been  shown to exhibit two distinct non-equilibrium steady states of uniform  (uNESS) and non-uniform  (nuNESS) particle distributions\cite{cheng21}.
The NESS refers to the situation that although particles flow in and out of each urn, but the average population remains constant in time in the steady state, and uNESS  corresponds to the case that the steady particle populations are  the same in each urn whereas nuNESS is for the case that the average steady populations in some of the urns are different.
As the inter-particle attraction varies, a first-order non-equilibrium phase transition occurs between these two NESSs characterized by a coexistence regime. The phase boundaries, the NESS particle distributions near the saddle points and the associated particle fluxes, average urn population fractions, and the relaxational dynamics to the NESSs were obtained analytically and verified numerically. A generalized non-equilibrium thermodynamics law  explicitly identifies the heat, work, energy, and entropy of the system was established.

In this paper, we report our investigations on the Ehrenfest urn model with interactions with an even number of urns placed on a ring and the  discovery of new possible non-equilibrium non-uniform steady states that are absent for an odd number of urns. In particular, we showed that   for four urns arranged in a ring, there is a new stable nuNESS phase  with minimal but non-vanishing non-uniformity in addition to the one with maximal non-uniformity which exists for any  $M$-urn on a ring  with $M\geq 3$. 
Our previous paper\cite{cheng21,cheng23} on the nonequilibrium behavior of urns in a ring in which the main explicit results were for $M=3$. The present  4-urn case is qualitatively different which is rooted in a different symmetry and the nuNESS emerges with a different bifurcation mechanism.
The paper is organized as follows, Sec. II gives a brief review of the multi-urn model at equilibrium and the notion of uniform and non-uniform population states. The major theoretical results are presented in Sec. III, including the complete phase diagram for the 4-urn ring systems, the signatures and generation mechanisms for various NESSs  together with the corresponding fluxes.
The  thermodynamic stability of these NESSs is analyzed in Sec. IV.  Sec. V presents the Monte Carlo simulations for the 3-urn and 4-urn ring models to verify explicitly  the validity of the theoretical results.

\section{Ehrenfest multi-urn  model with interaction: Equilibrium case}
The Ehrenfest multi-urn  model is based on the classic non-interacting Ehrenfest urn model with two urns but  generalized to $M$ urns  with the introduction of interactions for particles within the same urn\cite{cheng20}. For $N$ particles in $M$ urns in the large $N$ limit, 
the state of the system is labeled by the particle occupation fraction in each urn,
$\vec x=(x_1, x_2, \dots, x_{M-1})$ where $x_i$ is the fraction of particle  in the $i$-the urn, with $x_M=1-\sum_{i=1}^{M-1} x_i$. The energy of the interacting particles (in units of $k_BT\equiv 1/\beta$) in the urns is given by
\begin{equation}
\beta{\cal E}=\frac{Ng}{2}\sum_{i=1}^M x_i^2
\end{equation}
where $g<0$ denotes the pair-wise intra-urn particle attraction energy. 
For the equilibrium case, the jumping rates of a particle from one urn to another are the same, and detailed balance is obeyed. 
 The $(M-1)$-dimensional phase space is defined by the vector   ${\vec x}\equiv (x_1,\cdots,x_{M-1})^\intercal$.
The system can achieve thermal equilibrium and the equilibrium population distribution in the urns follows Boltzmann\cite{cheng20}, as
\begin{eqnarray}
\rho_{eqm}({\vec x})&=&{\cal N}\frac{e^{Nf({\vec x})}}{\sqrt{\prod_{i=1}^M x_i}}, \quad {\cal N}^{-1}\equiv \int_{\sum_{i=1}^{M-1}x_i\leqslant 1} \prod_{i=1}^{M-1}dx_i \frac{e^{Nf({\vec x})}}{\sqrt{\prod_{i=1}^M x_i}}\\
\text{where } f({\vec x})&=& -\sum_{i=1}^M (x_i\ln x_i+{g\over 2}x_i^2).\label{fx}\\
\end{eqnarray}

A particle in the $i$-th urn  jumps to
the $j$-th urn with the corresponding transition probability
$ T_{ij}\equiv   \frac{1}{{\rm e}^{-g(x_i-x_j)}+1}$.
Without interaction ($g=0$), we have $T_{ij} = \frac{1}{2}$.
 As the inter-particle interaction strength is varied,  phases of different levels of non-uniformity emerge and their stabilities are calculated analytically. Only the most non-uniform phase is stable and other non-uniform phases are all unstable at equilibrium\cite{cheng20}. In particular, the coexistence of locally stable uniform and the most non-uniform phases  connected by first-order transition occurs. The phase transition threshold and energy barrier were derived exactly together with the phase diagram  obtained analytically\cite{cheng20}. In addition, it was found that for even $M\geq 4$, a non-uniform state emerges at $g\leq -M$ but is always unstable and hence cannot be observed in practice.

\section{Non-equilibrium steady states for multi-urn  on a ring}
\begin{figure}[h]
  \begin{center}
    \includegraphics[width=3in]{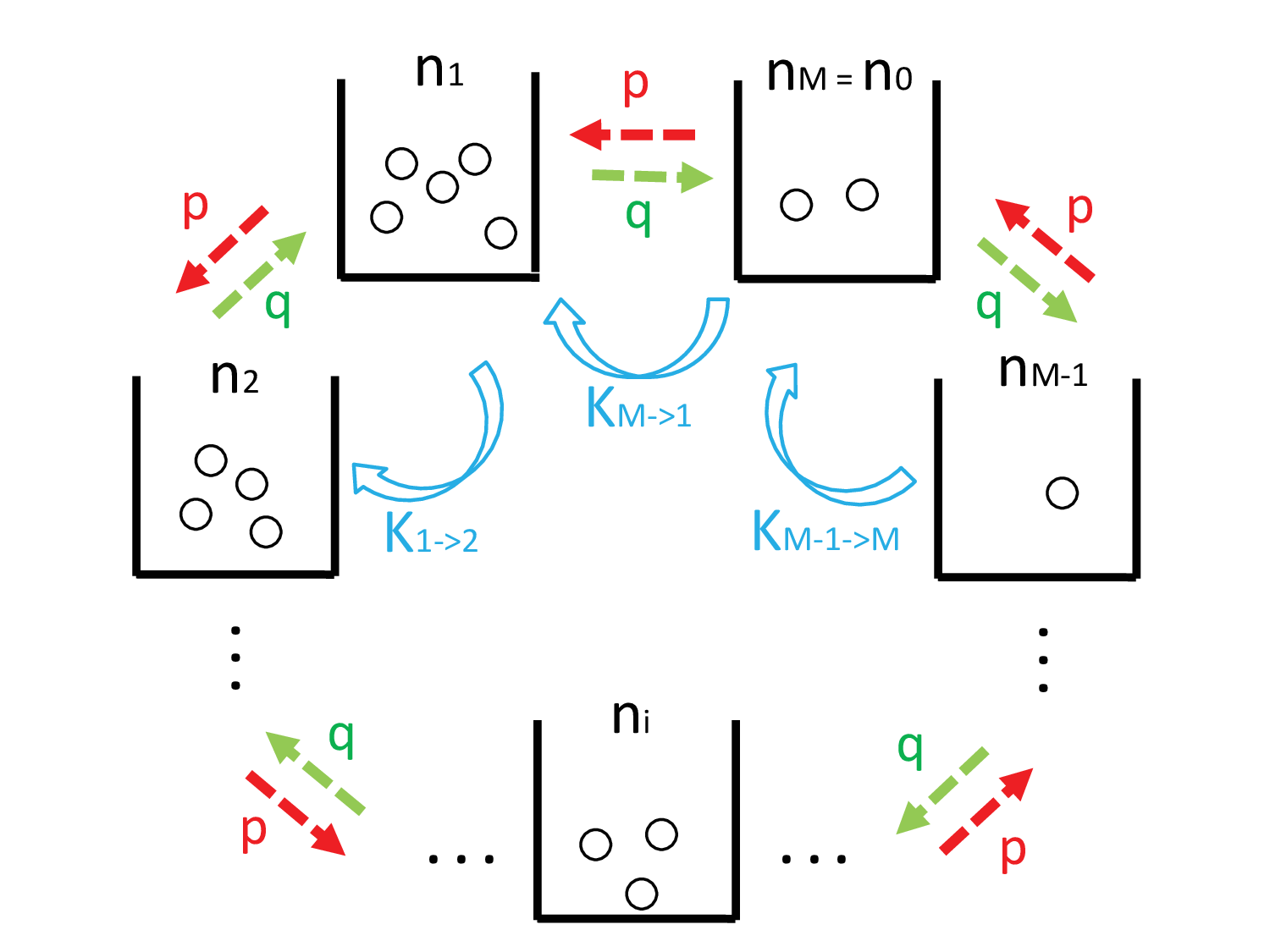}
  \end{center}
  \vspace{-5pt}
  \caption{Schematic picture of the interacting Ehrenfest $M$-urn  model on a ring. $M$ urns placed on  a ring and particle transitions are allowed between neighboring urns in.  The particle number in the $i$-th urn is denoted by
    $n_i$.  For convenience, we define $n_0 \equiv n_M$.
    The  jumping rates in the counter-clockwise and clockwise directions are $p$ and $q$ respectively. $K_{i\rightarrow j}$ represents the
    net particle flow rate from the $i$-th to the $j$-th urn.}
  \label{schematic.eps}
  \vspace{25pt}
\end{figure}

Non-equilibrium scenario occurs when there is a bias in the jumping rates of particles between the urns, and the non-equilibrium behavior depends on the connection network topology among the urns.
To be specific and for theoretical convenience, the system of
$M$ urns connected in a ring is considered as a paradigm model for investigating the non-equilibrium  steady-state properties and their associated thermodynamics.
The periodic boundary condition can be respected conveniently by defining the variable $x_0\equiv x_M$.
To establish the non-equilibrium states, a jumping rate is introduced such that the probability
of anticlockwise (clockwise) direction is $p$ ($q$).
For the sake of convenience, $p+q=1$ is imposed 
which only changes the time scale. $p=q=\frac{1}{2}$ reduces to the equilibrium scenario.
The schematic picture of the $M$-urn ring model was shown in \cite{cheng23} and is shown here again in Fig. \ref{schematic.eps} for completeness.
The population dynamics of the urns is governed by the following nonlinear coupled ODEs:
\begin{eqnarray}
\frac{d{\vec x}}{dt}&=&{\vec A}({\vec x}), \qquad\text{ where}\nonumber\\
A_i({\vec x})&\equiv& K_{i-1\to i}(x_{i-1}\to x_{i})-K_{i\to i+1}(x_{i},x_{i+1}), \qquad\text{ and}\label{Murns}\\
K_{i\to i+1}&=&\frac{p x_i -(1-p) x_{i+1} e^{g (x_{i+1}-x_i )}}{e^{g (x_{i+1}-x_i )}+1}\quad i=0,1,\cdots,M-1,\label{Kij}
\end{eqnarray}
is the net counter-clockwise particle flux from $i$ to $i+1$ urns.
 The corresponding $A_i(\vec x)$
are ($i=1,2,\cdots,M-1$)
\begin{widetext}
\begin{eqnarray}
  && A_i(\vec x) =
  - \frac{p x_i}{{\rm e}^{-g(x_i-x_{i+1})}+1}
  + \frac{q x_{i+1}}{{\rm e}^{-g(x_{i+1}-x_i)}+1}
  + \frac{p x_{i-1}}{{\rm e}^{-g(x_{i-1}-x_i)}+1}
  - \frac{q x_i}{{\rm e}^{-g(x_i-x_{i-1})}+1}    \label{Ai} 
\end{eqnarray}
\end{widetext}
which do not have explicit time dependence, i.e.  the system is autonomous.
To quantify how non-uniform the state is, one can  define
 \begin{equation}
    \Psi=\sqrt{\frac{1}{M(M-1)}\sum_{i\neq j}(x_i-x_j)^2}\label{psi}
\end{equation}
as the non-uniformity of the state\cite{cheng20}. $\Psi= 0$ for the uniform state and $\Psi$ is larger if the population fractions are more non-uniform.

The fixed points of the dynamical system (\ref{Murns}) are given by ${\vec A}({\vec x}^*)=0$, which leads to the condition of a constant flux between all urns:
\begin{equation}
K_{i\to i+1}=K_{ss} \qquad i=0,1,\cdots, M-1.\label{Kss}
\end{equation}
 It can be shown that  the equilibrium state is given by for $p={1\over 2}$  which is a necessary and sufficient condition for  $K_{i\to i+1}=0$. And   for $p\neq 1/2$,  a NESS with a non-zero constant $K_{ss}$ is possible.  
 The stability of the fixed point is  determined by the $(M-1) \times (M-1)$ Jacobian matrix ${\bf a}\equiv \frac{\partial {\vec A}}{\partial {\vec x}}|_{{\vec x}^*}$. The fixed point is dynamically stable if  there is no positive real part in all the eigenvalues of ${\bf a}$. In general, a potential cannot be derived with ${\vec A}({\vec x})=\nabla \Phi({\vec x})$ for some potential function $\Phi({\vec x})$ since the matrix ${\bf a}$ is in general asymmetric, even in the vicinity of the steady-state fixed point. For instance, at the uniform NESS fixed point, one still has ${\bf a}\neq {\bf a}^\intercal$ unless $p={1\over 2}$, in which equilibrium can be achieved and $\Phi={1\over 2}({\vec x}-{\vec x}^*)^\intercal {\bf a} ({\vec x}-{\vec x}^*)$ near the equilibrium fixed point.

\subsection{Phase diagrams for 3-urn and 4-urn models on  a ring }
\begin{figure}[H]
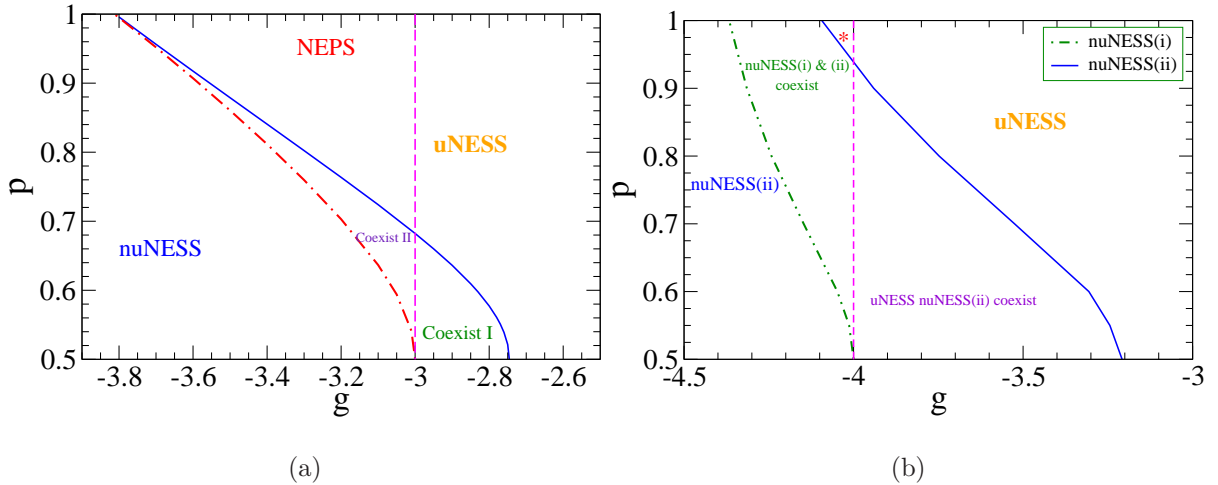

\centering
 \subfigure[]{\includegraphics*[width=.48\columnwidth]{phasediag3urns.eps}} 
  \subfigure[]{\includegraphics*[width=.48\columnwidth]{4urnsphasediag.eps}}
  \caption{(a) Phase diagram of the 3-urn  model  showing uNESS, nuNESS and NEPS.  The coexistence regions of  uNESS and nuNESS (coexist I),  nuNESS and NEPS (coexist II) are also labeled. (b) Phase diagram of the 4-urn  model showing different NESSs. The stability phase boundary obtained from (\ref{nuNESS1curve}) is shown by the dot-dashed curve. Only the nuNESS(i) phase exists in the region marked by {\color{red}*}.}\label{phasediagM4}
  \end{figure}

For discussion convenience, here we first show the phase diagrams obtained theoretically for the 3-urn and 4-urn models on a ring. The 3-urn phase diagram has been derived and discussed in detail in \cite{cheng21,cheng23}, and is shown here in Fig. \ref{phasediagM4}a  to compare with the phase diagram for the 4-urn case.
 The phase space is two-dimensional for the 3-urns model, and due to Poincar\'e Bendixon theorem\cite{strogatzbook}, the absence of a stable fixed point in some regime in the two-dimensional phase plane results in limit cycle oscillations. The transition from the uNESS  to NEPS at $g=-3$ is characterized by a supercritical Hopf bifurcation\cite{cheng23}, whereas the transition from the NEPS to nuNESS (the dot-dashed phase boundary in Fig. \ref{phasediagM4}a)  is characterized by an infinite-period bifurcation\cite{cheng23}. In addition, there are two coexisting regions:  Coexist I for the stable phases of uNESS and nuNESS, and Coexist II for NEPS and nuNESS. The presence of the coexistence regions signifies the corresponding first-order non-equilibrium phase transitions.
  For the new results on the 4-urn model on a ring, the phase diagram is displayed in Fig. \ref{phasediagM4}b showing the uniform NESS and two distinct non-uniform NESSs in which one of them (nuNESS(i)) cannot occur for odd values of $M$. The other nuNESS (nuNESS(ii)) corresponds to the NESS  that is maximally non-uniform, which is the same type as the nuNESS in the 3-urn model.  There are two non-overlap coexistence regions: the coexisting uNESS and nuNESS(ii) regime separated by the $g=-4$ line from the coexisting nuNESS(i) and nuNESS(ii) regime.
  The phase boundaries are derived and the  properties of these NESSs are discussed in the following subsections. 
  
\subsection{uNESS: NESS with the same population in each urn}
\begin{figure}[H]
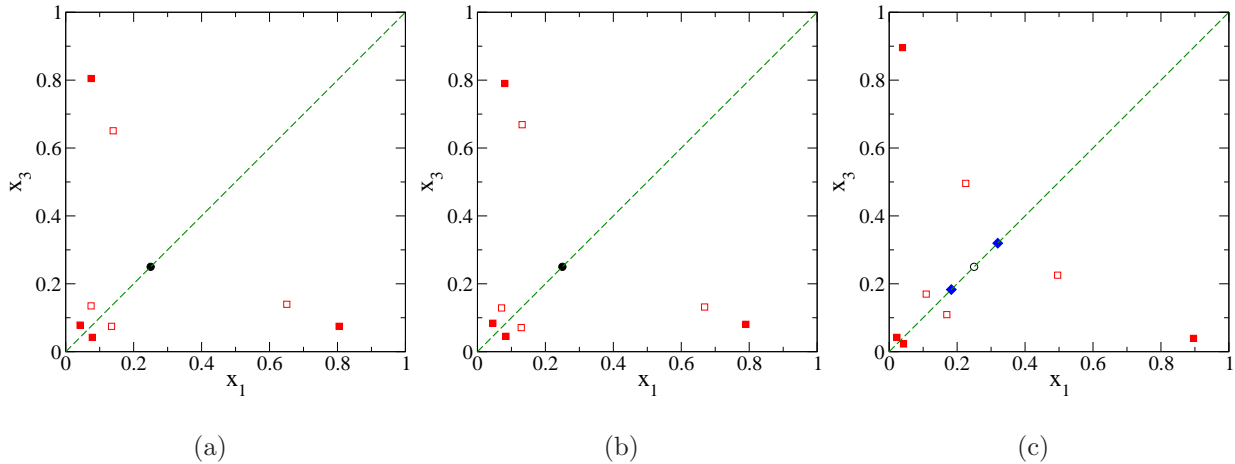

\centering
  \subfigure[]{\includegraphics*[width=.325\columnwidth]{nessp_8g-3_8.eps}} 
    \subfigure[]{\includegraphics*[width=.325\columnwidth]{nessp_8g-3_78.eps}} 
      \subfigure[]{\includegraphics*[width=.325\columnwidth]{nessp_8g-4_1.eps}}
  \caption{Projection of the phase space onto the $x_1-x_3$ plane showing the locations of the fixed points corresponding to different NESSs. Stable and unstable fixed points are denoted by filled and open symbols respectively. uNESS ($\circ$), nuNESS(i) ($\diamond$), nuNESS(ii) ($\square$). $p=0.8$.  (a) $g=-3.8$. (b) $g=-3.78$. (c) $g=-4.1$, the stable nuNESS(i) fixed point pair emerged.}\label{x1x3}
  \end{figure}

Notice that the uniform solution of $x_i^*=1/M$ is always a fixed point in (\ref{Murns}) with the flux
\begin{equation}
    K_{uNESS}=\frac{N(p-q)}{2M }=\frac{N(2p-1)}{2M }.
\end{equation}
The uNESS remains the only fixed point and is stable for $g>-M$, but becomes unstable for $g< -M$, as illustrated in  Fig. \ref{x1x3}.
Notice that the uNESS fixed point is independent of $p$ and $g$,  its stability is also independent of $p$.
In addition to the uniform state,  non-uniform fixed points with different values for $x_i^*$'s (non-uniform NESS) can exist in which some of them are related by symmetry. For example, Fig. \ref{x1x3}a and \ref{x1x3}b show the four symmetry-related stable and unstable non-uniform fixed point pairs  corresponds to nuNESS(ii) for the 4-urn ring.
  
 \subsection{nuNESS(ii): NESS with maximal non-uniformity--saddle-node bifurcation}
\begin{figure}[H]
\centering
\subfigure[]{\includegraphics*[width=.48\columnwidth]{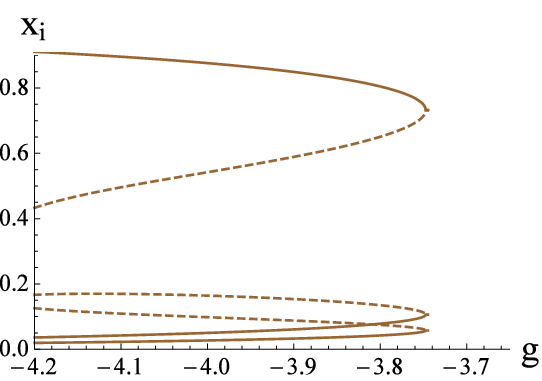}}
 \subfigure[]{\includegraphics*[width=.48\columnwidth]{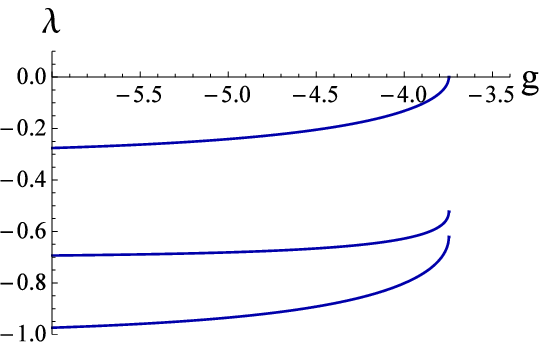}} 
  \caption{(a) Roots of the nuNESS(ii) fixed points as a function of $g$. The stable and unstable phases are denoted by solid and dashed curves respectively. (b) Eigenvalues plotted as function of $g$ with $p=0.8$   for the stable nuNESS(ii) phase in the 4-urn  model. }\label{xilam}
  \end{figure}

It has been shown in \cite{cheng21} for the 3-urn model that uNESS and nuNESS and their bistable co-existing states can occur. Such nuNESS (named nuNESS(ii) in this paper) occurs  via a saddle-node bifurcation in a regime of stronger attraction ($g$ sufficiently negative) and is characterized by the properties that one of the population fraction  is much larger than the rest (i.e. a state with maximal non-uniformity). For larger values of $M$, such nuNESS(ii) still persists. 
The emergence of the nuNESS(ii) can be understood by examining the non-trivial fixed points of (\ref{Murns}), which are plotted in Fig. \ref{xilam} for the 4-urn ring for illustration.
For large negative values of $g$, stable and unstable nuNESS(ii) states occur in pairs (filled and open squares in Fig. \ref{xs}).   As the inter-particle attraction decreases, the separation between the stable and unstable pair of roots decreases and annihilates each other at the phase boundary (solid curve in Fig. \ref{phasediagM4}).  Fig. \ref{xilam}a plots the stable (solid curves) and unstable (dashed curves) branches of the nuNESS(ii) fixed point pairs in the 4-urn ring as a function of $-g$. These fixed point pairs emerge via a saddle-node bifurcation as the inter-particle attraction increases to some threshold value. Fig. \ref{xilam}b shows the eigenvalues (which are all real and negative) of the stable saddle-node fixed point as a function of $g$, verifying its stability.
The phase boundary can be determined from the condition when the stable and unstable non-trivial fixed points of (\ref{Murns}) coincide, and are shown (solid curve) in the phase diagrams in Fig. \ref{phasediagM4}.

 \subsection{nuNESS(i): NESS with minimal but non-vanishing non-uniformity--pitchfork bifurcation for even number ($\geq 4$) of urn }
\begin{figure}[h]
\centering
 \subfigure[]{\includegraphics*[width=.325\columnwidth]{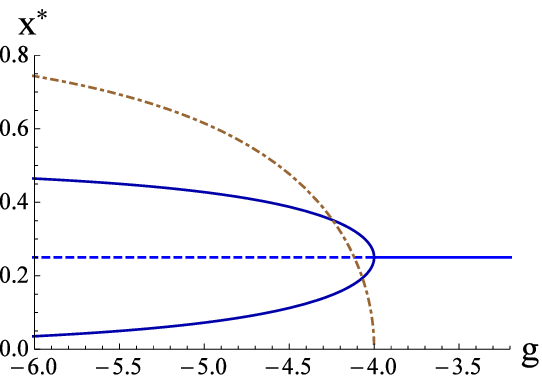}} 
   \subfigure[]{\includegraphics*[width=.325\columnwidth]{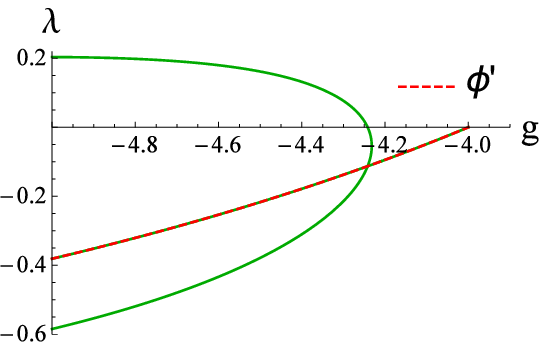}} 
   \subfigure[]{\includegraphics*[width=.325\columnwidth]{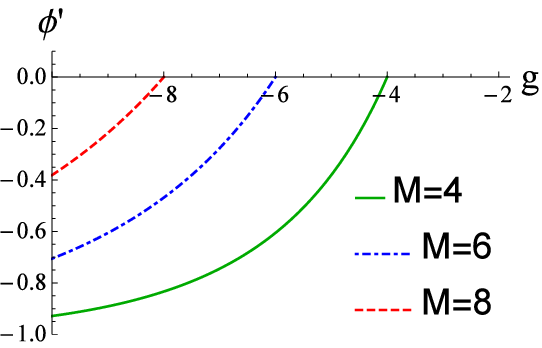}} 
  \caption{(a) Roots  from Eq. (\ref{rootnuNESS1}) for the nuNESS(i)  in  the 4-urn  model. The uNESS root of $\frac{1}{4}$  is shown by the horizontal line. A non-trivial symmetric pair of roots for the nuNESS(i) emerges when $g<-4$, indicating the classic scenario of a pitchfork bifurcation. The distance between the symmetric pair of roots in (three-dimensional) phase space as a function of $g$ is shown by the dot-dashed curve. (b) Real part of the eigenvalues of ${\bf a}$ plotted as function of $g$ with $p=0.8$ for the 4-urn  model. The eigenvalue  for the simplified model in Eq.  (\ref{model}) is also shown by the dashed curve, indicating that it is identical to the real branch of the eigenvalue of the nuNESS(i).(c)  Eigenvalue plotted as function of $g$ for the simplified model of the nuNESS(i). }\label{xs}
  \end{figure}

The nuNESS(i) is specified by the non-uniform population fraction that takes the form
$(x_1,x_2,\cdots,x_{M-1},x_{M})=( x^*,\frac{2}{M}- x^*, x^*,\frac{2}{M}- x^*,\cdots, x^*,\frac{2}{M}- x^*)$, where $ x^*$ can be determined from the root(s) of
\begin{equation}
    x^*=\frac{2}{M[1+e^{2g(x^*-\frac{1}{M})}]}.  \label{rootnuNESS1}
\end{equation}
$ x^*=\frac{1}{M}$ is always a trivial root that corresponds to uNESS, and we shall focus on the non-trivial root of $x^*\neq\frac{1}{M}$ and denote this root by $x_{pf}(g)$. 
 It is easy to see that $( x_{pf},\frac{2}{M}- x_{pf}, x_{pf},\frac{2}{M}- x_{pf},\cdots, x_{pf},\frac{2}{M}- x_{pf})$ is a root with minimal  but non-vanishing non-uniformity $\Psi$.
Notice that if $ x_{pf}$ is a non-trivial root, then so is $\frac{2}{M}- x_{pf}$, and hence the non-trivial roots always emerge in a symmetric (symmetric about $\frac{1}{M}$) pair.
The pair of fixed points are separated in phase space by a distance of $d=2\sqrt{M-1}\big|x_{pf}(g)-\frac{1}{M}\big|$.  Fig. \ref{x1x3}c shows the symmetric pair of stable nuNESS(i) fixed points (filled diamond) which lie on the $x_1=x_3$ line in the projected $x_1-x_3$ phase plane.   Fig. \ref{xs}a plots the roots in Eq. (\ref{rootnuNESS1}) as a function of $g$. $1/M$ is always a trivial uniform root that is independent of $g$ and becomes unstable when $g<-4$, accompanied by the emergence of a non-trivial symmetric stable pair of roots  for the nuNESS(i). The distance between the symmetric pair of roots in three-dimensional phase space  is also shown in Fig. \ref{x1x3}a.

The stability of nuNESS(i) can be revealed by examining the Jacobian matrix of this state for the 4-urn ring, ${\bf a}\equiv \frac{\partial {\vec A}}{\partial {\vec x}}|_{{x}_{pf}}$ can be calculated to give
\begin{eqnarray}
 {\bf a}&=&   \left(
\begin{array}{ccc}
{\scriptstyle  g (4-8 p) x^3+6 g p x^2-x (g p+g+2 p-2)-1} & {\scriptstyle (1-2 p) x \left(8 g x^2-6 g x+g+2\right)} &
 {\scriptstyle  -x \left(p \left(8 g x^2-6 g x+g+2\right)+2 g (1-2 x) x\right) }\\
{\scriptstyle (1-2 x) (p (g x (4 x-1)+1)+g (1-2 x) x)} &{\scriptstyle x (g (2 x-1)-2)} & {\scriptstyle(2 x-1) \left(p (g x (4
   x-1)+1)-2 g x^2-1\right)} \\
{\scriptstyle x (g (2 x-1) (p (4 x-1)-2 x+1)+2 (p-1)) }&{\scriptstyle (2 p-1) x \left(8 g x^2-6 g x+g+2\right)} &{\scriptstyle g
   (2 x-1) x (p (4 x-1)-2 x+2)+2 p x-1  } \\
\end{array}
\right)\Bigg|_{x_{pf}(g)}\label{anuNESS1}
\end{eqnarray}
whose eigenvalues can be explicitly computed as
\begin{eqnarray}
  & &  -1 - 2 g x_{pf}(g) (1 - 2 x_{pf}(g)),\quad \frac{1}{2} [ -1 - 2 g x (1 - 2 x) \pm\sqrt{\gamma(x,p,g)}]\Bigg|_{x_{pf}(g)}\\
{\scriptstyle \gamma(x,p,g) } &\equiv& {\scriptstyle 4 x^2 \left(g^2 (1-2 p)^2+10 g (1-2 p)^2+8 \left(2 p^2-2
   p+1\right)\right)+256 g^2 (1-2 p)^2 x^6-384 g^2 (1-2 p)^2 x^5}\\
   & &{\scriptstyle -4 x \left(g (1-2 p)^2+8
   p^2-8 p+4\right)+16 g (13 g+8) (1-2 p)^2 x^4-16 g (3 g+8) (1-2 p)^2 x^3+1}. \nonumber
\end{eqnarray}
Fig. \ref{xs}b plots the real part of the three eigenvalues as a function of $g$. For $g$ not much less than -4, there is only one real negative eigenvalue and the real part of the complex eigenvalue pair is also negative. As $g$ becomes more negative, all three eigenvalues become real and eventually one of the eigenvalues becomes positive and the nuNESS(i) loses its stability.
The stability phase boundary for the nuNESS(i) phase can be calculated theoretically from the condition of 
\begin{equation}
    -1 - 2 g x_{pf}(g) (1 - 2 x_{pf}(g)) +\sqrt{\gamma(x_{pf}(g),p,g)}=0,\label{nuNESS1curve}
    \end{equation}
    which gives the phase boundary $p=p_{pf}(g)$ as shown by the dot-dashed curve in the phase diagram  in Fig. \ref{phasediagM4}b. 

To analyze  the bifurcation nature of such nuNESS(i) phase for general (even) values of $M$ is challenging due to the dynamics in the high dimensional phase space.
To gain further insight of the nature of bifurcation at $g=-M$, notice that Eq. (\ref{rootnuNESS1}) can be rewritten as
\begin{equation}
    y=\frac{1}{M}\tanh[-g y],\quad y\equiv x_{pf}(g)-\frac{1}{M}.\label{rootb}
\end{equation}
Motivated by Eq. (\ref{rootb}), we propose the following one-dimensional simplified dynamical model to describe the bifurcation behavior of nuNESS(i):
\begin{equation}
    \dot{y}=\frac{1}{M}\tanh[-g y]-y\equiv \phi(y).\label{model}
\end{equation}
 It is easy to see that $y= 0$ is always a trivial fixed point in (\ref{model}), and a pair of  non-zero fixed points $\pm y^*\neq 0$, given by $\phi(y^*)=0$, emerge  for $g<-M$. The simplified model (\ref{model}) undergoes a classic supercritical pitchfork bifurcation in which the emerged $\pm y^*$ is always stable and accompanied by the loss of stability of the zero fixed point. The stability of $y^*$ can be verified  by calculating $\phi'(y^*)=-(1+g/M)+M {y^*}^2$, which is plotted as a function of $g$ for the 4-urn ring in Fig. \ref{xs}b. Remarkably, the eigenvalue of the simplified one-dimensional model (\ref{model}), $\phi'(y^*)$, is identical to one of the real eigenvalue of the full $M-1$-dimensional system, confirming the validity of the simplified model in the analysis of the pitchfork bifurcation of the nuNESS(i)  phase. For other even values of $M$, the eigenvalues  $\phi'(y^*)$ are always negative for $g<-M$, as shown in Fig. \ref{xs}c.

\subsection{ NESS fluxes and non-uniformity}
\begin{figure}[h]
\centering
\subfigure[]{\includegraphics*[width=.48\columnwidth]{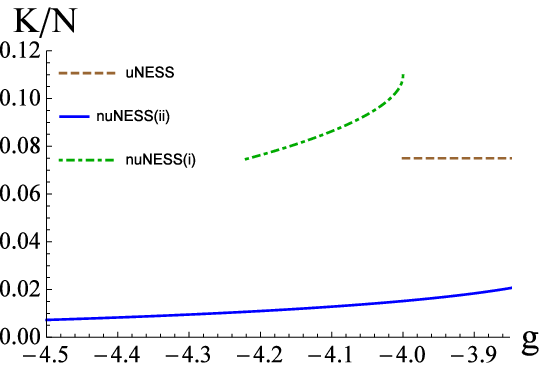}}
 \subfigure[]{\includegraphics*[width=.48\columnwidth]{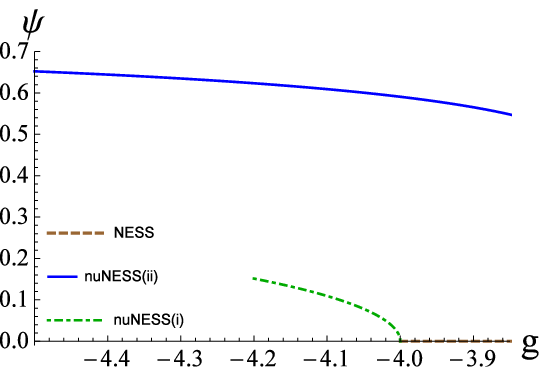}} 
  \caption{(a) NESS fluxes as a function of $g$ for $p=0.8$  for the uNESS, nuNESS(i) and nuNESS(ii) phases in the 4-urn ring.  (b) Non-uniformity $\Psi$ as a function of $g$  for the case in (a).}\label{flux}
  \end{figure}

The  NESS flux in the nuNESS(i) phase is given by
\begin{equation}
    K_{nuNESS(i)}=N\frac{(2p-1)x_{pf}(g)}{1+e^{g(\frac{2}{M}-x_{pf}(g))}}
\end{equation}
where $x_{pf}(g)$ is the root of $x$ in Eq. (\ref{rootnuNESS1}). The corresponding NESS flux  in this phase is evaluated from $K_{i\to i+1}$ in Eq. (\ref{Kij}) which can be shown to be independent of $i$ with $x_i$'s being the root of the nuNESS(i) stable fixed point.
Similarly, the non-uniformity is given by  (\ref{psi})  with the $x_i$'s taken to be the stable non-trivial fixed point. Fig. \ref{flux}a shows the NESS fluxes as a function of $g$ for fixed $p=0.8$, displaying the constant fluxes in various NESSs. In general, the nuNESS fluxes decrease as the attraction strength increases, and uNESS is significantly larger than nuNESS(ii). The nuNESS(i) flux is even greater than that of the uNESS.

The non-uniformity in the nuNESS(i) phase can be calculated by invoking (\ref{psi}) and is given by
\begin{equation}
    \Psi_{nuNESS(i)}=\sqrt{\frac{2M}{M-1}}\Big|x_{pf}(g)-\frac{1}{M}\Big|.
\end{equation}
 Fig. \ref{flux}b shows $\Psi$ in various NESSs as a function of $g$ for fixed $p=0.8$ for the 4-urn ring. $\Psi\equiv 0$ for uNESS as expected, and $\Psi$ increases with the inter-particle attraction strength. $\Psi$ is significantly larger in the nuNESS(ii) as compared with that of nuNESS(i), as anticipated. In addition, we verified for the 4-urn ring that nuNESS(i) is really the state with minimal  but non-vanishing non-uniformity and the stable nuNESS(ii) is really the state of maximal non-uniformity in the  sense that for all the roots in the fixed point equation ${\vec A}({\vec x})=0$ (all stable and unstable roots), the stable
nuNESS(i) and  nuNESS(ii)  phases are the states of minimal(but non-vanishing) and maximal values of $\Psi$ respectively.
  
\subsection{Absence of NEPS in the 4-urn ring}
As shown in \cite{cheng23}, NEPS occurs in the 3-urn model in the large $p$ and $g< -M$ regime. However, such a periodic non-equilibrium state does not exist for the 4-urn ring despite the emergence of complex eigenvalues pairs due to Hopf bifurcation for $g<-4$. This is due to the presence of the new stable nuNESS(i) phase  that attracts the otherwise periodic trajectories to this stable fixed point and kills the limit cycle. Since nuNESS(i) cannot exist for odd $M$, thus we anticipate stable oscillatory  dynamics such as NEPS  can occur for odd $M$.

\section{Fluctuations and Thermodynamic Stability of the NESSs}
Although a lot is known about   equilibrium fluctuations,
 current understanding of fluctuations in non-equilibrium
states is rather limited\cite{Ruelle}. For the  NESSs in our multi-urn system consisting of a large number of particles, there are considerable fluctuations in particle numbers and their fluxes.
These fluctuations may vary spatially in different urns
owing to the strong interactions between the particles and large fluctuations that can result from collective effects near the phase transitions.

The fluctuations in a NESS can be revealed by examining  the steady-state particle distribution functions. As shown in \cite{cheng21}, using the WKB (saddle-point) method, one can obtain the linearized Fokker-Planck equation and hence the steady-state distribution near the saddle-point can be described by the deviation ${\vec y}={\vec x}-{\vec x}^*$ as
\begin{equation}
\rho_{ss} \propto e^{N {\vec y}^\intercal {\bf c}  {\vec y}},\label{rhosssaddle} 
\end{equation}
where ${\vec x}^*$ is the stable fixed point of the steady state. The inverse of the $(M-1)\times (M-1)$ matrix ${\bf c}$ can be solved from the Lyapunov equation
\begin{equation}
{\bf a c}^{-1}+{\bf c}^{-1}{\bf a}^\intercal =2 {\bf b},\label{acb}
\end{equation}
with ${\bf a}_{ij}\equiv \partial_{x_j}A_i |_{{\vec x}^*} $ and ${\bf b}\equiv {\bf B}({\vec x}^*)$ with 
 $B_{ij}(\vec x)$ given by
 \begin{widetext}
\begin{eqnarray}
   && B_{ii}(\vec x) =
    \frac{p x_i}{{\rm e}^{-g(x_i-x_{i+1})+1}}
  + \frac{q x_{i+1}}{{\rm e}^{-g(x_{i+1}-x_i)}+1}
  + \frac{p x_{i-1}}{{\rm e}^{-g(x_{i-1}-x_i)}+1}
  + \frac{q x_i}{{\rm e}^{-g(x_i-x_{i-1})}+1}      \\
  && B_{i,i+1}(\vec x) = B_{i+1,i}(\vec x) =
  - \frac{p x_i}{{\rm e}^{-g(x_i-x_{i+1})}+1}
  - \frac{q x_{i+1}}{{\rm e}^{-g(x_{i+1}-x_i)}+1}.
\end{eqnarray}
\end{widetext}
It can be shown that the stability of the fixed point of the dynamical system implies that the  eigenvalues (which are real since ${\bf c}^{-1}$ is symmetric)  of $ {\bf c}^{-1}$ are all negative and hence the steady state is also thermodynamically stable\cite{cheng23}. The eigenvalues of ${\bf c}^{-1}$ can also provide valuable information on the local thermal fluctuating properties of the NESSs. Below we shall calculate $ {\bf c}^{-1}$ and examine its eigenvalues in various NESSs.
For the uNESS, the matrices ${\bf a}$,  ${\bf b}$ and hence ${\bf  c}^{-1}$ can be calculated explicitly for arbitrary values of $M$. For the 4-urn ring, we have for the uNESS:
\begin{eqnarray}
{\bf a}=-\frac{1}{2}
\left(
  \begin{array}{ccc}
    1+p+\frac{3g}{8} & 2p-1 &p+\frac{g}{8}\\
   - p-\frac{g}{8}& 1+\frac{g}{4} &-1+p-\frac{g}{8}\\
    1-p+\frac{g}{8}&1-2p&2-p+\frac{3g}{8}
  \end{array}
\right)
\end{eqnarray}
and its eigenvalues are  $-\frac{g+4}{4}$ and $-\frac{g+4}{8} \pm  (p-\frac{1}{2})i$. Notice that for NESSs with $p\neq \frac{1}{2}$, an imaginary part of the eigenvalues always exists, which gives rise to the oscillatory features in the NESSs.
${\bf b}$ and ${\bf  c}^{-1}$ for uNESS are given by
\begin{eqnarray}
{\bf b}=\frac{1}{8}
\left(
  \begin{array}{ccc}
    2&-1&0\\
   -1&2&-1\\
    0&-1&2
  \end{array}
\right),\quad\hbox{ and }
{\bf c}^{-1}=\frac{1}{g+4}
\left(
  \begin{array}{ccc}
   -3& 1 & 1 \\
    1 & -3&1 \\
    1&1&-3
  \end{array}
\right)
\end{eqnarray}
whose eigenvalues are $-\frac{1}{g+4}$ and $-\frac{4}{g+4}$ (multiplicity 2).
It can be seen that the real part of eigenvalues of $\bf a$ and the eigenvalues of ${\bf c}^{-1}$ are
always negative (positive) if $g>-4$ ($g<-4$), indicating that the uNESS becomes unstable both dynamically and thermodynamically as the attraction strength is beyond $-M$.

\begin{figure}[H]
\centering
\subfigure[]{\includegraphics*[width=.48\columnwidth]{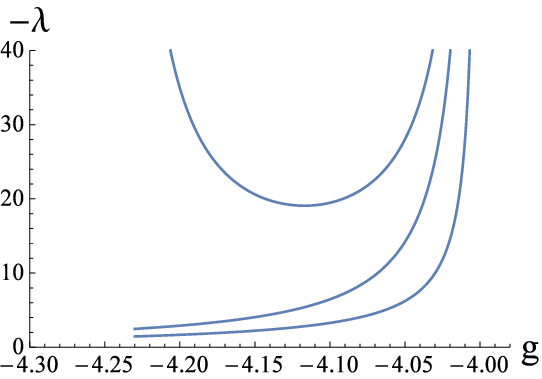}}
 \subfigure[]{\includegraphics*[width=.48\columnwidth]{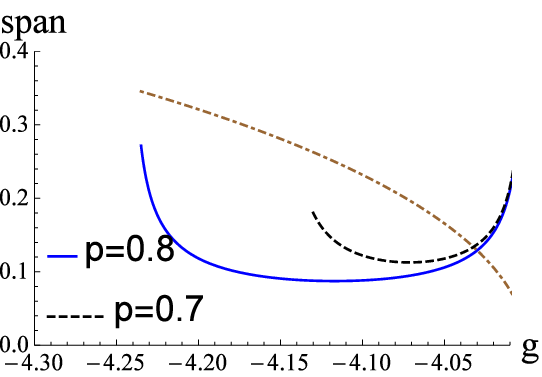}} 
  \caption{(a) Eigenvalues of  ${\bf c}^{-1}$ as a function of $g$ for $p=0.8$  for the nuNESS(i)  phase in the 4-urn ring.  (b) Span size of fluctuations  of the nuNESS(i), $2\sqrt{\max\{- \lambda_{{\bf c}^{-1}}\}/(2N)}$, as a function of $g$ for $p=0.7$ and 0.8. The distance between the symmetric pair of roots in phase space as a function of $g$ is shown by the dot-dashed curve.}
  \label{lambdacinv}
  \end{figure}

For nuNESS(i), ${\bf a}$ is given by (\ref{anuNESS1}) and 
\begin{eqnarray}
{\bf b}=[1-2x_{pf}(g)]x_{pf}(g)
\left(
  \begin{array}{ccc}
    2&-1&0\\
   -1&2&-1\\
    0&-1&2
  \end{array}
  \right).
\end{eqnarray}
${\bf c}^{-1}$ can then be solved from the Lyapunov equation (\ref{acb}).
Fig.\ref{lambdacinv}a shows the three eigenvalues of ${\bf c}^{-1}$, which are all negative, as a function of $g$ for a fixed value of $p=0.8$, verifying the thermodynamical stability of  nuNESS(i).

For nuNESS(ii), the explicit forms for ${\bf a}$ and ${\bf b}$ are tedious and will not be listed here. Nevertheless, they can be evaluated at the nuNESS(ii) fixed points to arbitrary accuracy, and ${\bf c}^{-1}$ can then be obtained from  (\ref{acb}).

\section{Monte Carlo Simulations}
To explicitly verify the theoretical results in previous sections, we carry out Monte Carlo simulations for the $M$ urns system. In the simulation, a total of $N$ ($N$ is an integer multiple of $M$) particles are in the system consisting of $M$ urns, and the population of the $i$ urn is denoted by $n_i$. 
The urns are placed on a bidirectional ring network with  anti-clockwise and clockwise  jump rates $p$ and $q$ respectively.  $p+q=1$ is imposed which only sets the time scale.  
The transition probability that  a particle from
the $i$th urn jumps to the $j$th urn is
\begin{equation}
T_{i\to j}=\frac{1}{1+e^{-\frac{g}{N}(n_i-n_j-1)}}.\label{Tij}
\end{equation}
A particle is chosen at random out of all the particles in the $M$ urns and a transition jump is made according to the probability given in (\ref{Tij}).
If $p=q$, then  with the above particle transition rules the system satisfies the detailed balance condition such that there is vanishing net particle flux on the ring. 
In general if $p>q$, there will be a net anti-clockwise flux and a NESS state can be achieved.
After some  sufficiently long transient time, the populations in each urn or the fraction $x_i(t)$ is recorded for a long sampling time.
Time is in Monte Carlo Steps per particle (MCS/N). One MCS/N means that on average every particle has attempted a jump.

\begin{figure}[H]
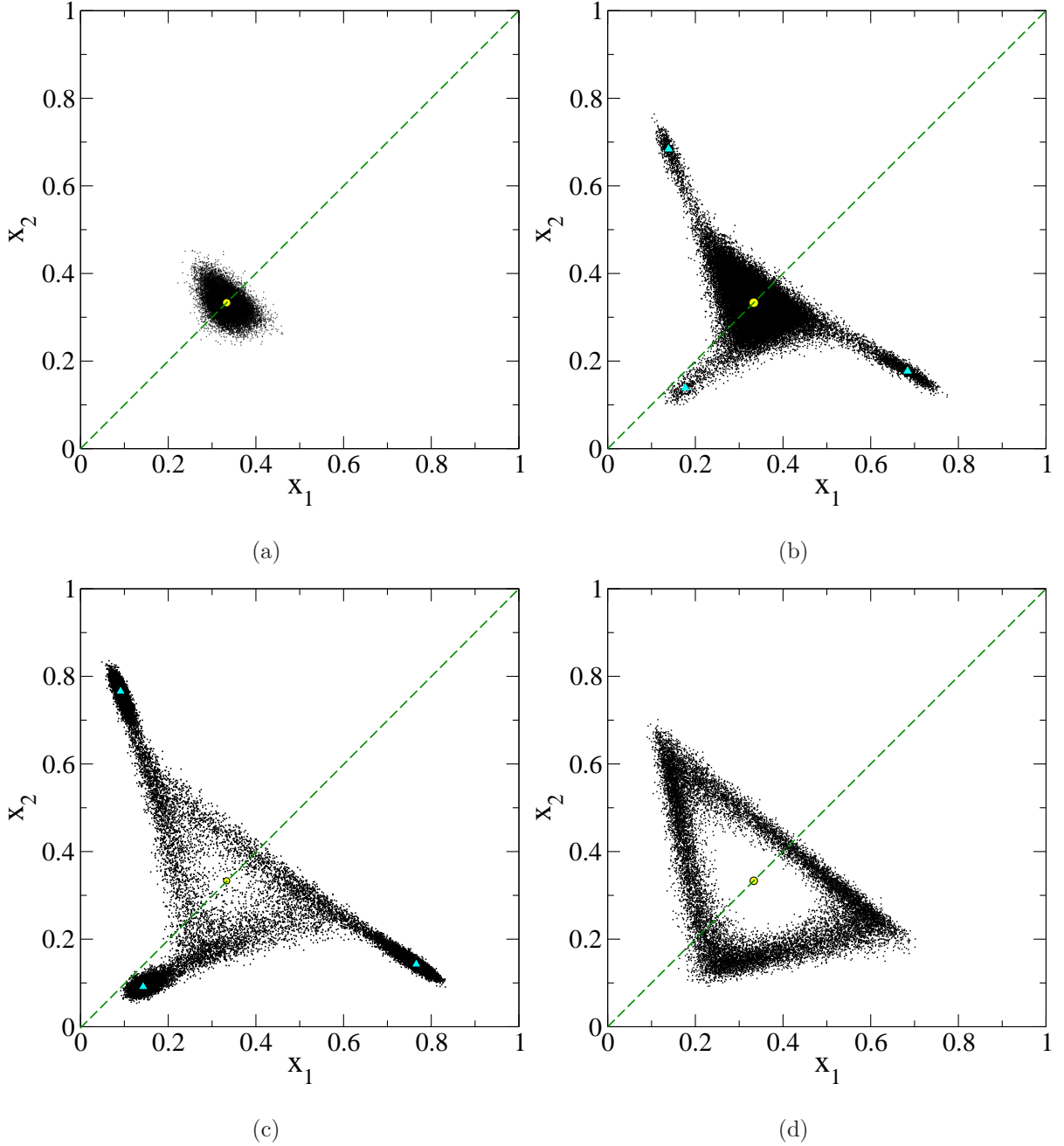

\centering
\subfigure[]{\includegraphics*[width=.48\columnwidth]{M3MCx1x2p_6g-2_6.eps}}
\subfigure[]{\includegraphics*[width=.48\columnwidth]{M3MCx1x2p_6g-2_85b.eps}}
\subfigure[]{\includegraphics*[width=.48\columnwidth]{M3MCx1x2p_7g-3_08b.eps}}
\subfigure[]{\includegraphics*[width=.48\columnwidth]{M3MCx1x2p_8g-3_1.eps}} 
 \caption{Monte Carlo simulation results of the population fraction map projected on the $x_1-x_2$ plane of 3000 particles in the 3-urn model. (a) $p=0.6$ and $g=-2.6$ in the uNESS. (b) $p=0.6$ and $g=-2.85$ in the uNESS and nuNESS coexistence region. (c) $p=0.7$ and $g=-3.08$ in the  nuNESS and NEPS coexistence region. (d) $p=0.8$ and $g=-3.1$ in the NEPS.} \label{M3map}
\end{figure}

Fig. \ref{M3map} plots the Monte Carlo simulation results of the population fraction map  for the 3-urn model showing various NESS, NEPS, and their coexistence regions as predicted 
 by the theory and shown in the phase diagram Fig. \ref{phasediagM4}a in previous section. For $g>-3$, the system stays around the uNESS and fluctuates around the stable $(\frac{1}{3}, \frac{1}{3})$ uniform fixed point as shown in Fig. \ref{M3map}a. The coexistence of uNESS and nuNESS (Coexist I) can be seen in Fig. \ref{M3map}b in which the system spends time around the stable uNESS and the three nuNESS saddle-points.  The Coexist II region with simultaneous occurrence of NEPS and nuNESS  can be seen in Fig. \ref{M3map}c showing the system switches from stable periodic NEPS to one of the nuNESS stochastically. Finally, the pure NEPS oscillation can be seen clearly in Fig. \ref{M3map}d in which the population fractions oscillate from high and low values periodically.

\begin{figure}[H]
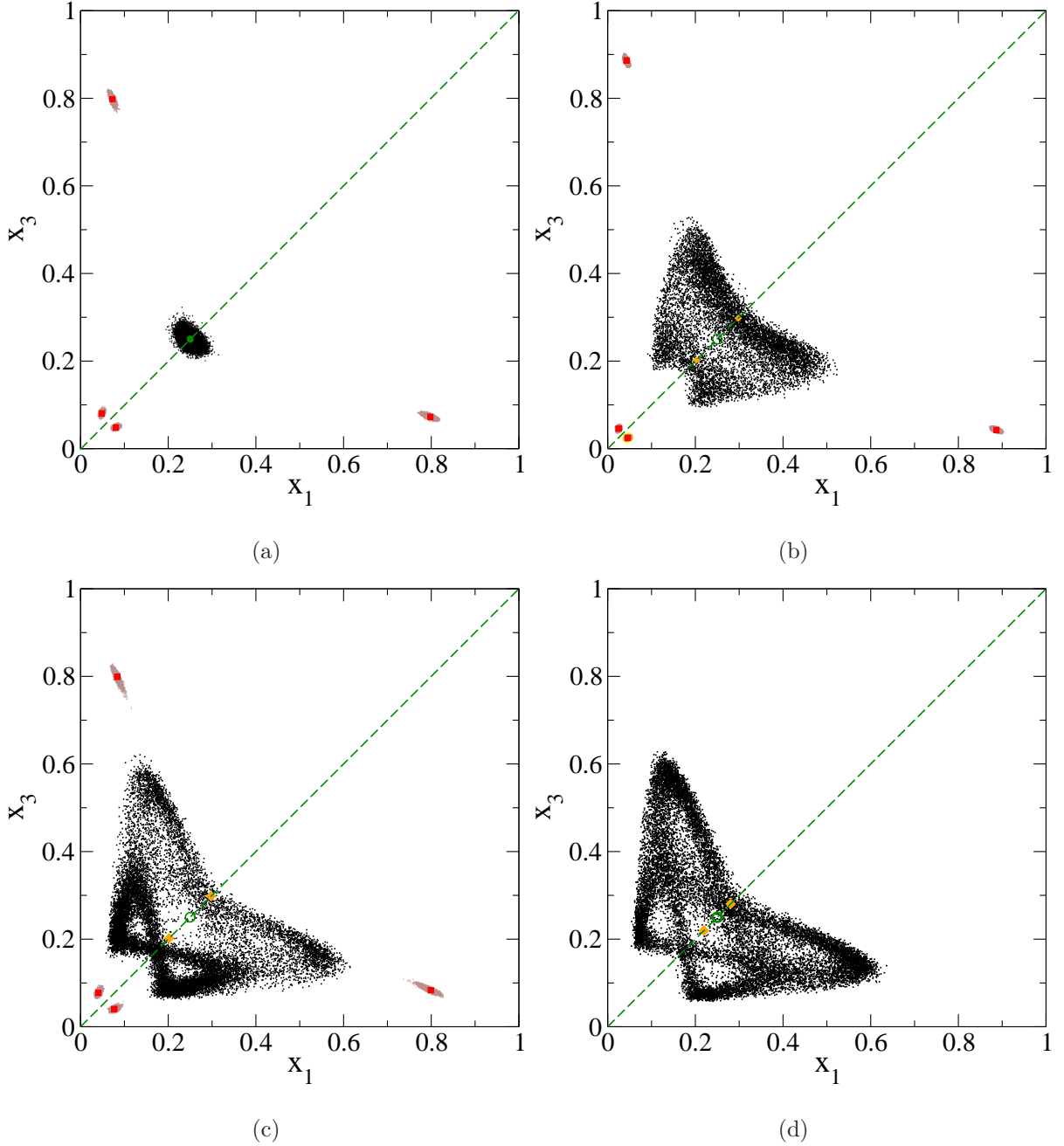

\centering
\subfigure[]{\includegraphics*[width=.48\columnwidth]{M4x1x3p_7g-3_6b.eps}}
\subfigure[]{\includegraphics*[width=.48\columnwidth]{M4x1x3p_8g-4_05.eps}}
\subfigure[]{\includegraphics*[width=.48\columnwidth]{M4x1x3p_95g-4_05b.eps}} 
\subfigure[]{\includegraphics*[width=.48\columnwidth]{M4x1x3p_975g-4_02.eps}} 
  \caption{Monte Carlo simulation result of the population fraction map projected on the $x_1-x_3$ plane of 20000 particles in the 4-urn model on a ring, for a simulation time of $8000$ MCS/N.  Each cluster of points starts from different initial conditions. (a) $p=0.7$ and $g=-3.6$ 
   showing the coexistence of the uNESS and nuNESS(ii) phases. The stable fixed points  are shown by the filled symbols:  uNESS  is marked by ($ \bullet$) and for nuNESS(ii) by ($\blacksquare$). (b) $p=0.8$ and $g=-4.05$ , showing the coexistence of the nuNESS(i) and nuNESS(ii) phases. The stable fixed points calculated from Eq. (\ref{rootnuNESS1}) for nuNESS(i) ($\blacklozenge$) and    nuNESS(ii) ($\blacksquare$) are shown by the filled symbols. The unstable uNESS fixed point of $(\frac{1}{4},\frac{1}{4})$ is also marked by an open circle ($\circ$). (c)  $p=0.95$ and $g=-4.05$ , showing the coexistence of the nuNESS(i) and nuNESS(ii) phases. (d) Only the nuNESS(i) exists for $p=0.975$ and $g=-4.02$. }\label{MCx1x3}
  \end{figure}

MC simulations are also carried out for the 4-urn ring to confirm the new nuNESS(i) in the previous section. Fig. \ref{MCx1x3} plots the projection of the population fraction map onto the $x_1-x_3$ plane to show the various NESSs and their coexistence regimes.
The coexistence of uNESS and nuNESS(ii) is shown in Fig. \ref{MCx1x3}a, in which the parameters are chosen such that these NESSs are far apart in phase space and transition from one NESS to another is quite impossible in the affordable duration of simulation times.
The coexistence of the two nuNESSs is shown in Fig. \ref{MCx1x3}b. Notice that the stochastic fluctuations around two symmetric nuNESS(i) saddle-points (marked by filled circles along the $x_1=x_3$ line) give rise to the smear basin of the nuNESS(i) phase characterized by a relatively improbable region near the unstable uNESS saddle-point (open circle). When the parameters for the coexistence of nuNESS(i) and nuNESS(ii) are closer to the phase boundary, the basin of the nuNESS(i) phase appears to be broadened, as shown in Fig. \ref{MCx1x3}c. Finally, only pure nuNESS(i) is observed for high values of $p$ and $g\lesssim -4$ (the * region in the phase diagram Fig. \ref{phasediagM4}b) as shown in Fig. \ref{MCx1x3}d. Stable NEPS is never observed in the 4-urn ring simulations.

To understand the large spreading of the population fraction for the nuNESS(i) state, one can estimate the width of the distribution about the nuNESS(i) saddle-point, which can be obtained from the eigenvalues of ${\bf  c}^{-1}$ as $2\sqrt{\max\{- \lambda_{{\bf c}^{-1}}\}/(2N)}$. Fig. \ref{lambdacinv}b plots the span of the fluctuations about the nuNESS(i) fixed points, indicating the fluctuations become very large near the phase boundaries. Furthermore, since there are two symmetric nuNESS(i) saddle-points separated by a distance (shown by the dot-dashed curve in  Fig. \ref{lambdacinv}b) that is comparable with the fluctuation spans, the two symmetric stable nuNESS(i) saddle-points have large overlaps in stochastic fluctuations and thus resulted in a rather smear basin of attraction of the stochastic trajectories as observed in Fig. \ref{MCx1x3}b-\ref{MCx1x3}d.

\section{Summary }
In this paper, the Ehrenfest urn model with interactions with an even number of urns placed on a ring is investigated for new possible non-equilibrium non-uniform steady states. For the 4-urn ring, we proved that indeed there exists a new stable nuNESS phase (nuNESS(i)) with minimal (but non-vanishing) non-uniformity in addition to the one with maximal non-uniformity  (nuNESS(ii)) and the uNESS that were reported for the 3-urn case. Such a new nuNESS(i) phase emerged from a pitchfork bifurcation that is only possible for an even number of urns. There are two coexistence regions, one for the coexisting uNESS and nuNESS(i) and another for coexisting nuNESS(i) and nuNESS(ii). The phase diagram together with the phase boundaries for the 4-urn ring and their NESS fluxes are calculated theoretically. These findings are also confirmed by explicit Monte Carlo simulations of the 3-urn and 4-urn ring models.
In addition, the physics due to the distinct features of symmetric pair emerged nuNESSs allows one to investigate the characteristics of the non-equilibrium phase transitions between uNESS and nuNESS, as well as between distinct nuNESSs not related by symmetry.

Compared with the 3-urn case, the 4-urn model can allow the coexistence of  distinct nuNESSs, not related by symmetry, which correspond to the minimal and maximal non-uniformity. Since the internal entropy production rate is a decreasing function of the non-uniformity, it implies that the traditional principle of minimum or maximum entropy production is invalid in the selection of the most stable non-uniform NESS among them. 

Higher values of $M>4$ will be investigated in the future by simulations and by stability analysis in detail since more complex dynamics might result due to the higher dimensionality of the phase space involved.
Since the $M$-urn ring model has recently been shown\cite{cheng23b} to be related to the $M$-state Potts model\cite{Potts} with special dynamical rule, our result indicates possible new non-equilibrium states in the even number of state Potts model.
\bibliographystyle{unsrt}
\bibliography{references}
\end{document}